\documentclass{SCIS2016}
\usepackage{hyperref}

\begin{document}

\ArticleType{RESEARCH PAPER}
{\put(0,-26){5G Wireless Communication Networks}}
\Year{2016}
\Month{January}
\Vol{59}
\No{1}
\DOI{xxxxxxxxxxxxxx}
\ArtNo{xxxxxx}
\ReceiveDate{January 1, 2016}
\AcceptDate{January 1, 2016}
\OnlineDate{January 1, 2016}

\title{5G green cellular networks considering power allocation schemes}{5G green cellular networks considering power allocation schemes}

\author[1]{GE Xiaohu}{}
\author[1]{CHEN Jiaqi}{}
\author[2]{WANG Chengxiang}{}
\author[3]{THOMPSON John}{}
\author[1]{ZHANG Jing}{{zhangjing@mail.hust.edu.cn}}

\AuthorMark{GE Xiaohu}

\AuthorCitation{Ge Xiaohu, Chen Jiaqi, Wang Chengxiang, et al}

\address[1]{School of Electronics Information and Communications, Huazhong University of Science and Technology,\\
 Wuhan {\rm 430074}, Hubei, P. R.China}
\address[2]{Joint Research Institute for Signal and Image Processing, School of Engineering \& Physical Sciences,\\
 Heriot-Watt University, Edinburgh {\rm EH14 4AS}, UK}
\address[3]{Institute for Digital Communications, University of Edinburgh, Edinburgh {\rm EH9 3JL}, UK}
\maketitle

\abstract{It is important to assess the effect of transmit power allocation schemes on the energy consumption on random cellular networks. The energy efficiency of 5G green cellular networks with average and water-filling power allocation schemes is studied in this paper. Based on the proposed interference and achievable rate model, an energy efficiency model is proposed for MIMO random cellular networks. Furthermore, the energy efficiency with average and water-filling power allocation schemes are presented, respectively. Numerical results indicate that the maximum limits of energy efficiency are always there for MIMO random cellular networks with different intensity ratios of mobile stations (MSs) to base stations (BSs) and channel conditions. Compared with the average power allocation scheme, the water-filling scheme is shown to improve the energy efficiency of MIMO random cellular networks when channel state information (CSI) is attainable for both transmitters and receivers.}

\keywords{Energy efficiency, cellular networks, MIMO, achievable rate model, power allocation scheme}

\citationline

\section{Introduction}
A more than ten-fold increase in mobile data traffic between 2013 and 2018 is predicted in recent forecasts from Ericsson and Cisco~\cite{Cisco14}. Corresponding to this growth rate in mobile communications, 15--20\% energy consumption for the entire information and communications technologies (ICT) industry, as well as 0.3--0.4\% of annual global carbon dioxide emissions will be increased~\cite{Chen10}. Considering the significant proportion of mobile data traffic, it is important to more deeply analyze the energy efficiency of 5G green cellular networks and provide some guidelines for future power allocation scheme in the fifth generation (5G) cellular networks.

With the development of wireless transmission technologies, multi-input multi-output (MIMO) antenna technology is widely used to improve the capacity of wireless communication systems. Moreover, numerous energy efficiency models have been investigated for MIMO communication systems in~\cite{Alfano12,liu12, jiang14, chen13, jiang13, jiang12, Garcia11, chong1, joung13}. To maximize the energy efficiency of MIMO communication systems over time varying channels, the impact of line-of-sight, out-of-cell interferers and the antenna correlation was discussed for downlink channels in~\cite{Alfano12}. An optimal power control algorithm was proposed for the generalized energy-efficiency proportional fair metric in a multiuser MIMO communication system~\cite{liu12}. A tight upper bound of the energy efficiency with a spectrum efficiency constraint was derived for a virtual-MIMO communication system which has one destination and one relay using the compress-and-forward (CF) cooperation scheme~\cite{jiang14}. Based on the proposed energy efficiency upper bound, the optimal power and bandwidth allocation have been derived for maximizing the energy efficiency of MIMO communication systems. In~\cite{chen13}, an energy efficient adaptive transmission scheme was proposed for MIMO beamforming communication systems with orthogonal space-time block coding (OSTBC) with imperfect channel state information (CSI) at transmitters. An algorithm that jointly considering the transmit power, power allocation among streams and beamforming matrices was developed to maximize the energy efficiency of MIMO communication systems with interference channels~\cite{jiang13}. Due to the trade-off between the traffic rate and the hardware power consumption, an antenna selection algorithm was developed in MIMO communication systems~\cite{jiang12}. By jointly choosing the transmission power and precoding vector among codebooks, a radio resource optimization scheme was proposed to improve the spectrum and energy efficiency of MIMO communication systems with user fairness constraints~\cite{Garcia11}. Assuming that channel state information is known to the transmitters, an optimal power control scheme was proposed for maximizing the energy efficiency of a base station (BS) using multiple antennas \cite{chong1}. Considering distributed transmitter systems employing a zero-forcing based multiuser MIMO precoding, a heuristic power control method was proposed to improve the energy efficiency of MIMO communication systems under constraints on the per-user target rate and the per-antenna instantaneous transmit power~\cite{joung13}. However, the above studies concerning the energy efficiency of MIMO communication systems have been limited to finite numbers of interfering transmitters.

Many studies indicated that improving the energy efficiency of cellular networks is a critical problem for the future of the telecommunication industry~\cite{Davaslioglu14, Hasan11, zou13, han13, Li14, Nguyen12, Soh13, Karray10, xiang13}. The purpose of traditional cellular wireless communications always is higher throughput for the user and higher capacity for the service provider, regardless of energy efficiency. Davaslioglu {\em et. al.} discussed the specific reasons for inefficiency and potential improvement in the physical layer as well as in more higher layers of the communication protocol of cellular networks~\cite{Davaslioglu14}. Hasan {\em et. al.} presented a review of methods of improving the energy efficiency of cellular networks, and explored some related topics and challenges, moreover suggested some techniques to make green cellular networks possible~\cite{Hasan11}. A novel user cooperation scheme termed inter-network cooperation was investigated to improve uplink emission energy efficiency of cellular networks with the help of a short-range communication network~\cite{zou13}. Three typical multi-cell cooperation scenarios, i.e., the energy efficiency coordinated multi-point transmissions, the traffic-intensity-aware and the energy-aware multi-cell cooperation were also discussed for reducing the energy consumption of cellular networks in~\cite{han13}. The downlink performance evaluation of small cell networks including capacity and energy efficiency was investigated in~\cite{Li14}, where BSs and users are modeled as two independent spatial Poisson point processes. In related work on a MIMO cellular network with one single macrocell base station (MBS) and multiple femtocell access points, an opportunistic interference alignment scheme was proposed for reducing the intra and inter tier interference and the energy consumption~\cite{Nguyen12}. Through the deployment of sleeping strategies and small cells, the success probability and energy efficiency were improved for homogeneous macrocell single tier wireless networks and heterogeneous multiple tiers wireless networks in~\cite{Soh13}. Using the signal-to-interference-and-noise ratio (SINR) as the function of the user's location, an analytical model was proposed for calculating the spectrum and energy efficiency of cellular networks with orthogonal frequency division multiplexing access (OFDMA)~\cite{Karray10}. Based on single antenna transmission systems, the energy efficiency of random cellular networks with the statistical analysis of traffic load and power consumption was also evaluated in~\cite{xiang13}.

However, in future 5G mobile communication systems, the energy efficiency is proposed as one of the most important performance indicators~\cite{Chen151, Chen152, Chen153}. Considering that the 5G network will be a huge multi-layer multi-RAT Het-Net network~\cite{Wang14}, simple scenarios such as MIMO communication systems considering finite interfering transmitters in one single cell are so simple that have no ability to accurately evaluate the energy efficiency of complex cellular networks. Moreover, studies of the impact of different power allocation schemes, which is the important influence factor in power consumption evaluation, on the energy efficiency of MIMO random cellular networks are surprisingly rare in the open literature. Motivated by the previous issues, we investigate the energy efficiency of MIMO random cellular networks with different power allocation schemes in this paper. The contributions and novelties are summed up as follows.
\begin{itemize}
\itemindent 2.8em
\item[(1)] A random cellular network using stochastic geometry theory for MIMO transmitters and receivers is proposed for evaluating the network level energy efficiency considering different power allocation schemes.
\item[(2)] We propose an interference model and compute the achievable rate of MIMO PVT random cellular networks with infinite interfering BSs which distributing as a Poisson point process, taking effects of path loss, fading and shadowing in radio propagation channels into account.
\item[(3)] With the proposed interference and achievable rate model, performance analysis for energy efficiency of MIMO PVT random cellular networks with average and water-filling power allocation schemes have been derived.
\item[(4)] Based on the simulation results, the impact of average and water-filling power allocation schemes on the energy efficiency of MIMO PVT random cellular networks has been analyzed and compared in detail.
\end{itemize}

\section{System model}
Assume that in the infinite plane ${{{\mathbb{R}}}^2}$ BSs and MSs are deployed randomly, of which the locations are approximated to be two independent Poisson point processes~\cite{Yang03} with intensities ${\lambda _M}$ and ${\lambda _B}$, which are expressed as
\begin{equation}
{\Pi _B} = \left\{ {{y_{Bi}},i = 0,1,2, \cdots } \right\},{\Pi _M} = \left\{ {{x_{Mj}},j = 0,1,2, \cdots } \right\},
\tag{1}
\end{equation}
where ${y_{Bi}}$ and ${x_{Mj}}$ are two-dimensional location coordinates of the $i$th BS $B{S_{i}}$ and the $j$th MS $M{S_{j}}$, respectively.

\begin{figure}
\centering
\includegraphics[width=9cm,draft=false]{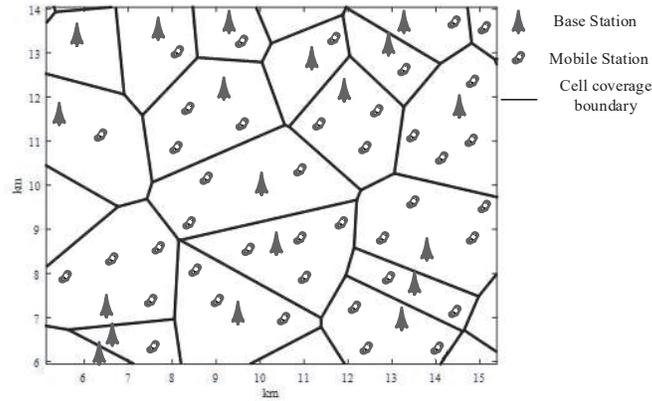}
\caption{ Illustration of PVT random cellular networks.\label{F1}}
\end{figure}

Assume that MSs communicate with the closest BS for suffering the minimum path loss in the process of radio propagation. All other BSs in the infinite plane ${{{\mathbb{R}}}^2}$ are interfering BSs. The OFDMA scheme is adopted for wireless transmission to avoid the intra-interference in the cell. Thus, we can split the plane ${{{\mathbb{R}}}^2}$ into a number of irregular polygons approximately expressing coverage areas of different cells through the Delaunay Ttiangulation method~\cite{Feren07}. The illustration of stochastic and irregular topology in Figure~\ref{F1} is so-called Poisson Voronoi Tessellation (PVT) random network, where each cell is identified by ${{C_i}\left( {i = 0,1,2, \cdots } \right)}$.
\\[1mm]

According to Palm theory, one of most important features of PVT random cellular networks is that geometric characteristics of all cells coincide with each other, such that can be viewed as coinciding with a typical PVT cell ${C_0}$~\cite{Stoyan96}. Thus, analytical results for a typical PVT cell ${C_0}$ can reveal properties of the whole PVT random cellular network.

Assume that each BS is integrated with ${N_t}$ transmission antennas and each MS is equipped with ${N_r}$ receive antennas. In this paper, our study is focused on the downlinks of cellular wireless communication systems. Without loss of generality, the signal received at an MS $M{S_0}$ in the typical PVT cell ${C_0}$ is expressed as
\[\begin{gathered}
  {{\bm{y}}_0} = {{\bm{H}}_{00}}{{\bm{x}}_0} + \sum\limits_{i = 1}^\infty  {{{\bm{H}}_{i0}}{{\bm{x}}_i}}  + {{\bm{n}}_0} \hfill \\
  \quad\;  = {{\bm{H}}_{00}}{{{\bm{\hat V}}}_0}{{\bm{\varLambda }}_0}{{\bm{c}}_0} + \sum\limits_{i = 1}^\infty  {{{\bm{H}}_{i0}}{{{\bm{\hat V}}}_i}{{\bm{\varLambda }}_i}{{\bm{c}}_i}}  + {{\bm{n}}_0} \hfill \\
\label{2}
\end{gathered} ,\tag{2}\]
where ${{\bm{c}}_0}$ is ${N_t}$ dimension desired transmitted symbol satisfying ${\bm{c}}_0^H{{\bm{c}}_0} = 1$ from the BS $B{S_0}$, ${{\bm{c}}_i}\left( {i = 1,2, \cdots } \right)$ is the interfering transmitted symbol satisfying ${\bm{c}}_i^H{{\bm{c}}_i} = 1$ from the interfering BS $B{S_i}\left( {i = 1,2, \cdots } \right)$, ${{\bm{\varLambda}}_i}= diag\left( {\sqrt {{P_{i1}}} ,\sqrt {{P_{i2}}} , \ldots ,\sqrt {{P_{i{N_t}}}} } \right)$ $\left( {i = 0,1,2, \cdots } \right)$ is ${N_t}\times {N_t}$ transmit power allocation vector satisfying $\sum\limits_{j = 1}^{{N_t}} {{P_{ij}}}  = {P_{Ti}}$. The scalars ${P_{T0}}$ and ${P_{Ti}}\left( {i = 1,2, \cdots }\right)$ denote the transmission power at the desired BS $B{S_0}$ and the interfering BS $B{S_i}$ respectively. Afterwards, the transmitted symbol vector is precoded by an ${N_t} \times {N_t}$ matrix ${{{\bm{\hat V}}}_i}$ as ${{\bm{x}}_i}={{{{\bm{\hat V}}}_i}{{\bm{\varLambda }}_i}{{\bm{c}}_i}}\left( {i = 0,1,2, \cdots } \right)$; ${{\bm{H}}_{00}}$ is the ${N_r}\times{N_t}$ channel matrix between the MS $M{S_0}$ and the desired BS $B{S_0}$, ${{\bm{H}}_{i0}}(i = 1,2,3 \ldots )$ is the ${N_r}\times{N_t}$ channel matrix between the MS $M{S_0}$ and the interfering BS $B{S_i}$, the element ${h_{0,k,n}}(k = 1,2, \ldots ,{N_r};n = 1,2, \ldots ,{N_t})$ of channel matrix ${{\bm{H}}_{00}}$ and the element ${h_{i,k,n}}(i = 1,2, \ldots ;k = 1,2, \ldots ,{N_r};n = 1,2, \ldots ,{N_t})$ of channel matrix ${{\bm{H}}_{i0}}$ are independently and identically distributed (\textit{i.i.d.})\footnotemark[1]; ${\bm{n}_0}$ is the ${N_r}$ dimension additive white Gaussian noise (AWGN) vector in the wireless channel, the noise power is equal to ${N_0}$. Due to the infinite sum of interferers in equation~\eqref{2}, it is reasonable to assume that the system model of MIMO PVT random cellular networks is an interference limited scenario.

\footnotetext[1]{Our analysis can also approximate fading correlation scenarios by performing moment matching to simplify to a single Gamma distribution~\cite{Ahmadi09}.}

\section{Achievable Rate of MIMO PVT Random Cellular Networks}

\subsection{Interference Model}
The received signals including the interference signals are assumed to be propagated though independent wireless channels~\cite{Kostic05}. The shadowing effect is assumed to be follow a log-normal distribution, to which a Gamma fading distribution is an alternative approximately~\cite{Bithas06} for simplifying calculation. The multi-path fading is assumed to be follow a Nakagami-m distribution which spans via the m parameter the widest range of the amount of fading (from 0 to 2) among all the multi-path distributions~\cite{Simon00}. In this case, the wireless channel gain from the $n{\text{th}}$  transmission antenna at the interfering BS $B{S_i}$ to the $k{\text{th}}$ receive antenna at the MS $M{S_0}$ is expressed as
\begin{equation}
{\left| {\left| {{h_{i,k,n}}} \right|} \right|^2} = \frac{1}{{R_i^\sigma }}{w_{i,k,n}}{\left| {{z_{i,k,n}}} \right|^2},
\tag{3}
\end{equation}
where ${R_i}$ is the Euclidean distance between the MS $M{S_0}$ and the interfering BS $B{S_i}$, $\sigma$ is the path loss coefficient in radio propagation,  ${w_{i,k,n}}$ is a random variable governed by Gamma distribution, and ${z_{i,k{\text{,}}n}}$ is a random variable governed by Nakagami-m distribution.

Considering that the OFDMA scheme and a relevant interference cancellation scheme~\cite{Dai12} are used for intra-cell signals, there is no significant co-channel interference from within one PVT cell~\cite{Annapureddy11}. Therefore, the co-channel interference is assumed to be transmitted from all BSs in the infinite plane except for the BS in typical PVT cell ${C_0}$. Assume that the active interfering BSs set is  modeled an independent thinning process on the BS Poisson point process, which still form a Poisson point process with intensity ${\lambda _{{\text{Inf}}}}$~\cite{xiang13}, and generally satisfies $0 \leqslant {\lambda _{{\text{Inf}}}} \leqslant {\lambda _B}$. Therefore, the interference power aggregated at the MS  $M{S_0}$ is expressed as~\cite{Win09}
\begin{equation}
{P_I} = \sum\limits_{k = 1}^{{N_r}} {\left( {\sum\limits_{i = 1}^\infty  {\frac{{{I_{i,k}}}}{{R_i^\sigma }}} } \right)}  = \sum\limits_{i = 1}^\infty  {\frac{{{I_i}}}{{R_i^\sigma }}}  = \sum\limits_{i \in {\Pi _{\inf }}} {\frac{{{I_i}}}{{R_i^\sigma }}},
\tag{4a}
\label{4a}
\end{equation}
with
\begin{equation}
{I_i} = \sum\limits_{k = 1}^{{N_r}} {\left( {{I_{i,k}}} \right)} ,
\tag{4b}
\end{equation}
where ${I_{i,k}}$ is the received interference of the $k{\text{th}}$ antenna of the MS $M{S_0}$  from the  $i{\text{th}}$  BS regardless of the pass loss fading. Considering that every antenna of the MS $M{S_0}$  will receive multiple interference streams transmitted from ${N_t}$ antennas of interfering BSs,  ${I_i}$ represents the interference power received by ${N_r}$  antennas at the MS $M{S_0}$, which is further expressed as~\cite{Ge11}
\begin{equation}
{I_i} = {P_{Ti}}\sum\limits_{k = 1}^{{N_r}} {\sum\limits_{n = 1}^{{N_t}} {{T_{i,k,n}}} } ,
\tag{5a}
\end{equation}
with
\begin{equation}
{T_{i,k,n}} = {w_{i,k,n}}{\left| {{z_{i,k,n}}} \right|^2}.
\tag{5b}
\end{equation}
Assume that the average power of interference terms that are transmitted from the $n$th antenna of interfering BS $B{S_i}$  and received at the $k$th antenna of MS $M{S_0}$  are approximately equal in the statistical meaning~\cite{Alouini99}. Thus, the Gamma fading over all sub-channels is simplified as ${w_{i,k,n}}(i = 1,2, \ldots ;k = 1,2, \ldots ,{N_r};n = 1,2, \ldots ,{N_t}) = {w_i}$. Furthermore, the interference power received by ${N_r}$  antennas of the MS $M{S_0}$  is derived by
\begin{equation}
{I_i} = {P_{Ti}}{H_i},
\tag{6a}
\end{equation}
with
\begin{equation}
{H_i} = {w_i}\sum\limits_{k = 1}^{{N_r}} {\sum\limits_{n = 1}^{{N_t}} {{{\left| {{z_{i,k,n}}} \right|}^2}} } .
\tag{6b}
\end{equation}

A channel that experiences the product of both Gamma fading and Nakagami fading follows a ${K_G}$  distribution. Therefore, the PDF of  ${H_i}$ is derived by~\cite{Bithas06}~\cite{Abdi98}~\cite{Gradshteyn07}
\[{f_{{H_i}}}\left( y \right) = \frac{{2{{\left( {\frac{{m\lambda }}{\Omega }} \right)}^{\frac{{{N_t}{N_r}m + \lambda }}{2}}}}}{{\Gamma \left( {{N_t}{N_r}m} \right)\Gamma \left( \lambda  \right)}}{y^{\frac{{{N_t}{N_r}m + \lambda  - 2}}{2}}} {K_{\lambda  - {N_t}{N_r}m}}\left( {2\sqrt {\frac{{m\lambda y}}{\Omega }} } \right)\left( {y > 0,i = 1,2,3...} \right),\tag{7a}\label{7a}\]
with\[\Omega {\text{ = }}\sqrt {\left( {\lambda {\text{ + }}1} \right)/\lambda } \;\;\;\lambda  = 1/{\left( {{e^{{{\left( {{\sigma _{dB}}/8.686} \right)}^2}}} - 1} \right)^2},\tag{7b}\]
where $\Gamma \left( \cdot \right)$ is a Gamma function, $m$ is a Nakagami shaping factor, ${K_{\lambda  - m}}\left( \cdot \right)$  is the modified Bessel function of the second kind with order $\lambda  - m$ and ${\sigma _{dB}}$  is the variance of shadowing effect values.

Based on equation~\eqref{4a} and the Campbell theory in~\cite{Feren07}, the characteristic function of the interference power aggregated at the MS $M{S_0}$  can be written as
\[\begin{gathered}
  {\Phi _{{P_I}}} = \mathbb{E}\left\{ {{e^{j\omega {P_I}}}} \right\}
   = \exp \left( { - 2\pi {\lambda _{\inf }}\iint {\left( {1 - {e^{\frac{{j\omega y}}{{{R^\sigma }}}}}} \right){f_I}\left( y \right)dyRdR}} \right) \hfill \\
   = {\text{exp}}\left( { - 2\pi {\lambda _{\inf }}\int_r {\left[ {1 - {\phi _I}\left( {\frac{\omega }{{{R^\sigma }}}} \right)} \right]RdR} } \right) \hfill \\
\end{gathered} ,\tag{8}\]
where  ${f_I}(y)$ and ${\phi _I}\left( \omega  \right)$  are the PDF and the characteristic function of the total interference power ${I_i}$  received at the MS $M{S_0}$, respectively; $\mathbb{E}\left\{\cdot\right\}$ is the expectation operation. Based on the result in~\cite{Win09}, the characteristic function ${\Phi _{{P_I}}}$ represents a alpha stable random process, which can be simply denoted as ${P_I} \sim \textrm{Stable}\left( {\alpha  = 2/\sigma ,\beta  = 1,\delta ,\mu  = 0} \right)$, where $\alpha$ and $\delta$ are the stability parameter and the scale parameter, respectively. Based on the alpha stable characteristic function expression, ${\Phi _{{P_I}}}$ can be re-written as
\[{\Phi _{{P_I}}}{=}\exp \left\{ {{-}\delta {{\left| \omega  \right|}^\alpha }\left[ {1{-}j\beta \textrm{sign}\left( \omega  \right)\tan \left( {\frac{{\pi \alpha }}{2}} \right)} \right]} \right\},\tag{9a}\]
with\[\begin{gathered}
  \delta  = {\lambda _{\inf }}\frac{{\pi \Gamma \left( {2 - \alpha } \right)\cos \left( {\frac{{\pi \alpha }}{2}} \right)}}{{1 - \alpha }}\mathbb{E}\left\{ {P_{Ti}^\alpha } \right\}\mathbb{E}\left\{ {H_i^\alpha } \right\}
  \left( {\alpha  \ne 1,i = 1,2,3 \ldots } \right) \hfill \\
\end{gathered} ,\tag{9b}\label{9b}\]
where $\mathbb{E}\left\{ {P_{Ti}^\alpha } \right\}$ is the moment of the receiving power raised to the power $\alpha$ at the MS $M{S_0}$. Based on equation~\eqref{7a}, $\mathbb{E}\left\{ {H_i^\alpha } \right\}$  is expressed as
\[\begin{gathered}
  \mathbb{E}\left\{ {H_i^\alpha } \right\} = \int\limits_0^\infty  {\frac{{2{{\left( {\frac{{m\lambda }}{\Omega }} \right)}^{\frac{{{N_t}{N_r}m + \lambda }}{2}}}}}{{\Gamma \left( {{N_t}{N_r}m} \right)\Gamma \left( \lambda  \right)}}{y^{\frac{{{N_t}{N_r}m + \lambda  - 2}}{2}}}}
{K_{\lambda  - {N_t}{N_r}m}}\left( {2\sqrt {\frac{{m\lambda y}}{\Omega }} } \right){y^\alpha }dy \hfill \\
\end{gathered}  .\tag{10}\label{10}\]
From the table of integrals in \cite{Gradshteyn07}, equation~\eqref{10} can be written in closed form as
\[\begin{gathered}
  \mathbb{E}\left\{ {H_i^\alpha } \right\} = {\left( {\frac{{m\lambda }}{\Omega }} \right)^{ - \alpha }}\frac{{\Gamma \left( {\lambda  + \alpha } \right)\Gamma \left( {{N_t}{N_r}m + \alpha } \right)}}{{\Gamma \left( {{N_t}{N_r}m} \right)\Gamma \left( \lambda  \right)}}
  \left( {i = 0,1,2,3 \ldots } \right) \hfill \\
\end{gathered} .\tag{11}\label{11}\]
Substituting equation~\eqref{11} into~\eqref{9b}, the PDF of the interference power aggregated at the MS $M{S_0}$  can be written as
\[{f_{{P_I}}}\left( y \right) = \frac{1}{{2\pi }}\int\limits_{ - \infty }^{ + \infty } {{\Phi _{{P_I}}}\left( {j\omega } \right)\exp \left( { - 2\pi j\omega y} \right)d\omega } .\tag{12}\label{12} \]

\subsection{Achievable Rate Model}
Based on the proposed interference model of MIMO random cellular networks in (12), the achievable rate at the MS is derived in this section. We assume the network is interference rather than noise limited, due to the infinite sum of interferers in equation~\eqref{2} ~\cite{Ge11}. Therefore, the received signal-to-interference ratio (SIR) at the MS $M{S_0}$ in the typical PVT cell ${C_0}$  is expressed as
\[\begin{gathered}
SI{R_0} = \frac{{{\bm{c}}_0^H{\bm{\varLambda }}_0^H{\bm{\hat V}}_0^H{\bm{H}}_{00}^H{{\bm{H}}_{00}}{{{\bm{\hat V}}}_0}{{\bm{\varLambda }}_0}{{\bm{c}}_0}}}{{{P_I}}}
\end{gathered} ,\tag{13}\]
where $H$ is the conjugate transpose operation. Furthermore, the achievable rate at the MS $M{S_0}$ is expressed as
\[\begin{gathered}
{{\cal R}_0} = {B_W}\log \left[ {1 + SI{R_0}} \right] = {B_W}\log \left[ {1 + \frac{{{\bm{c}}_0^H\bm{\varLambda}_0^H{\bm{\hat V}}_0^H{\bm{H}}_{00}^H{{\bm{H}}_{00}}{{{\bm{\hat V}}}_0}{\bm{\varLambda}_0}{{\bm{c}}_0}}}{{{P_I}}}} \right]
 \end{gathered} ,\tag{14}\]
where  ${B_W}$ is the bandwidth allocated for the MS $M{S_0}$ .

Transmitters are assumed to obtain the CSI from receivers without delay via uplink feedback channels. Moreover, the MIMO channel is divided into a number of  parallel single-input single-output (SISO) channels by the single value decomposi-tion (SVD) method. In this case, the channel matrix ${{\bm{H}}_{{{00}}}}$  in equation~\eqref{2} can be decomposed as
\[{{\bm{H}}_{{{00}}}}{{ = }}{{\bm{U}}_{00}}{{\bm{D}}_{00}}{\bm{V}}_{00}^H ,\tag{15}\]
where ${{\bm{D}}_{00}} = \textrm{diag}\left( {\sqrt {{\lambda _1}} ,\sqrt {{\lambda _2}} ,...,\sqrt {{\lambda _L}} }\right)$ is the ${L} \times {L}$ diagonal matrix, ${\lambda _1} \le {\lambda _2} \le ... \le {\lambda _L}$, are eigenvalues of the matrix  ${\bm{H}}_{_{00}}^H{{\bm{H}}_{00}}$, and $L={\textrm{rank}{\rm{(}}{{\bm{H}}_{00}}{\rm{)}}}$ is the rank of ${{\bm{H}}_{00}}$, ${{\bm{U}}_{00}}$ is the ${N_r} \times {L}$  unitary matrix, ${{\bm{V}}_{00}}$  is the ${L} \times {N_t}$  unitary matrix. Furthermore, assuming that the full matrix ${{\bm{V}}_{00}}={{{\bm{\hat V}}}_0}$, the achievable rate at the MS  $M{S_0}$ is expressed as
\[{{\cal R}_0} = {B_W}\log \left[ {1 + \frac{{{\bm{c}}_0^H{\bm{\varLambda }}_0^H{\bm{D}}_{00}^2{{\bm{\varLambda }}_0}{{\bm{c}}_0}}}{{{P_I}}}} \right] = {B_W}\log \left[ {1 + \frac{{{\bm{\varLambda }}_0^H{\bm{D}}_{00}^2{{\bm{\varLambda }}_0}}}{{{P_I}}}} \right].\tag{16}\label{16}\]

\section{Green MIMO Random Cellular Networks}
Considering that the MS required traffic load will influence the BS transmission power, the energy efficiency of MIMO PVT cellular networks will be related with the traffic rates in MSs. In this section, we discuss this relationship in more detail. Two classical power control schemes are discussed with the proposed model, numerical results show inherent relationships among the energy efficiency, traffic load, and the prevailing channel environment conditions.

\subsection{Basic Energy Efficiency Model}
In this paper, we define the energy efficiency of MIMO PVT cellular networks as the average ratio of traffic load over total power consumption at a BS in a typical PVT cell ${C_0}$~\cite{P.Chong11} based on Palm theory~\cite{Feren07}
\[EE = \frac{{r\left( P \right)}}{{P{{\left( P \right)}_s}}} = \frac{\mathbb{E}{\left\{ {{\varGamma _{{C_0}}}} \right\}}}{\mathbb{E}{\left\{ {{P_{BS}}} \right\}}}\left[ {\frac{{{\rm{nat}}}}{{{\rm{Joule}}}}} \right],\tag{17}\]
where ${\varGamma _{{C_0}}}$ is the traffic rate in a typical PVT cell ${C_0}$, ${P_{BS}}$ is the total power consumed at a BS in a typical PVT cell ${C_0}$. The total BS power consumption includes both fixed power and dynamic power consumption terms~\cite{Arnold10}. According to~\cite{Yu11}, the total power consumption ${P_{BS}}$ is written following
\[{P_{BS}} = \frac{{{P_{{C_0}\_real}}}}{\eta } + {N_t}{P_{dyn}} + {P_{sta}},\tag{18}\label{18}\]
where ${P_{{C_0}\_real}}$ is the total BS radio frequency transmission power for all transmit antennas, $\eta $ is the average efficiency of the BS power amplifiers, ${N_t}$ is the number of active BS antennas, ${P_{dyn}}$ is the RF circuit power for an antenna and ${P_{sta}}$ is the fixed power consumption in a BS. Moreover, there is a maximal BS transmission power ${P_{\max }}$ in practical. In this case, when the required BS transmission power exceeds ${P_{\max }}$ the corresponding transmission traffic will be interrupted. Therefore, the average transmission traffic rate is calculated by $\mathbb{E}{\left\{ {{\varGamma _{{C_0}}}} \right\}}{F_{{P_{{C_0}}}}}\left( {{P_{\max }}} \right)$, where ${F_{{P_{{C_0}}}}}\left( {{P_{\max }}} \right)$ is the probability that the BS transmission power less than ${P_{\max }}$. Furthermore, the energy efficiency of the MIMO PVT random cellular networks is expressed as
\[\begin{gathered}
  EE = \frac{{\mathbb{E}\left\{ {{\varGamma _{{C_0}}}} \right\}{F_{{P_{{C_0}}}}}\left( {{P_{{C_0}\_\max }}} \right)}}{{\mathbb{E}\left\{ {{P_{BS}}} \right\}}}
= \frac{{\mathbb{E}\left\{ {{\varGamma _{{C_0}}}} \right\}{F_{{P_{{C_0}}}}}\left( {{P_{{C_0}\_\max }}} \right)}}{{\frac{1}{\eta }\mathbb{E}\left\{ {{P_{{C_0}\_real}}} \right\} + {N_T}{P_{dyn}} + {P_{sta}}}} \hfill \\
\end{gathered} ,\tag{19}\label{19}\]
where $\mathbb{E}\left\{ {{P_{{C_0}\_real}}} \right\}$ is the average BS actual transmission power in the typical PVT cell ${C_0}$.

Many empirical measurement results have demonstrated that the traffic load in both wired and wireless networks, including cellular networks, is self-similar and bursty. Considering the infinite variance characteristic of self-similar distributions, Pareto distributions with infinite variance were proposed for modeling the self-similar traffic in wireless networks~\cite{Silva02}. Therefore, the traffic rate $\rho \left( {{x_{Mi}}} \right)$ at the MS $M{S_i}$ is assumed to be a Pareto distribution with infinite variance. Traffic rates of all MSs are assumed to be \textit{i.i.d.}. Then, the PDF of traffic rate is expressed by
\[{f_\rho }(x) = \frac{{\theta \rho _{\min }^\theta }}{{{x^{\theta  + 1}}}},{\rm{    }}x \ge {\rho _{\min }} > 0,\tag{20}\label{20}\]
where $\theta  \in \left( {1,2} \right]$ is a shape parameter, also known as  the tail index. ${\rho _{\min }}$  is minimum possible value of traffic rate that is needed to meet the MS's quality of service (QoS) requirements. Furthermore, the average traffic rate at a MS is expressed as
\[\mathbb{E}\left\{ \rho  \right\} = \frac{{\theta {\rho _{\min }}}}{{\theta  - 1}}.\tag{21}\]

Based on the results in \cite{Kostic05}, the average traffic rate for all MSs in a typical PVT cell ${C_0}$ is denoted as $\mathbb{E}\left\{ {{\varGamma _{{\cal C}0}}} \right\} = \frac{{{\lambda _M}\theta {\rho _{\min }}}}{{{\lambda _B}(\theta  - 1)}}$. As a consequence, the energy efficiency of MIMO PVT random cellular networks is derived by
\[EE = \frac{{\frac{{{\lambda _M}\theta {\rho _{\min }}}}{{{\lambda _B}(\theta  - 1)}}{F_{{P_{{C_0}}}}}\left( {{P_{{C_0}\_\max }}} \right)}}{{\frac{1}{\eta }\mathbb{E}\left\{ {{P_{{C_0}\_real}}} \right\} + {N_T}{P_{dyn}} + {P_{sta}}}}.\tag{22}\label{22}\]

\subsection{Energy Efficiency of MIMO Cellular Networks Using Average Power Allocation Scheme}
The average power allocation scheme is a simple antenna power control scheme which has been widely used for practical MIMO wireless communications systems. The maximum ratio transmission / maximum ratio combining (MRT/MRC) methods are assumed to be adopted in MIMO PVT random cellular networks ~\cite{Dighe03}, the achievable rate with an average power allocation scheme satisfying ${{\bm{\varLambda }}_0} = \frac{{{P_{0}}}}{{{N_t}}}{\bm{I}}\left( {{N_t}} \right)$ at the MS $M{S_0}$ in the typical PVT cell ${C_0}$ is derived as
\[\begin{gathered}
  {\cal{R}_0} = {B_W}\log \left[ {1{\text{  +  }}\frac{{{\bm{\varLambda }}_0^H{\bm{D}}_{00}^2{{\bm{\varLambda }}_0}}}{{{P_I}}}} \right]\; \hfill \\
  \;\;\; \;\;\;\leqslant {B_W}{\text{log}}\left[ {1{\text{ + }}\frac{{\frac{{{P_0}}}{{{N_t}}}{\lambda _{\max }}({\bm{H}}_{00}^H{{\bm{H}}_{00}})}}{{{P_I}}}} \right] {\text{ = }}{B_W}{\text{log}}\left[ {1{\text{ + }}\frac{{\frac{{{P_0}}}{{{N_t}}}{\lambda _{rank({{\bm{H}}_{00}})}}}}{{{P_I}}}} \right]\; \hfill \\
  \;\;\; \;\;\;\leqslant {B_W}\log \left[ {1{\text{ + }}\frac{{\frac{{{P_0}}}{{{N_t}}}\left\| {{{\bm{H}}_{00}}} \right\|_F^2}}{{{P_I}}}} \right] {\text{ = }}{B_W}{\text{log}}\left[ {{\text{1 + }}\frac{{\frac{{{P_0}}}{{{N_t}}}\sum\limits_{k{\text{ = }}1}^{{N_r}} {\sum\limits_{{\text{n = }}1}^{{N_t}} {{{\left| {{h_{0,k,n}}} \right|}^2}} } }}{{{P_I}}}} \right] \hfill \\
  \;\;\;\;\;\; \approx {B_W}{\text{log}}\left[ {{\text{1 + }}\frac{{\frac{{{P_0}}}{{{N_t}}}\sum\limits_{k{\text{ = }}1}^{{N_r}} {\sum\limits_{{\text{n = }}1}^{{N_t}} {\frac{1}{{R_{\text{0}}^\sigma }}{w_0}{{\left| {{z_{0,k,n}}} \right|}^2}} } }}{{{P_I}}}} \right] {\text{ = }}{B_W}{\text{log}}\left[ {{\text{1 + }}\frac{{\frac{{{P_0}}}{{{N_t}}}\frac{{{H_{\text{0}}}}}{{{\text{R}}_0^\sigma }}}}{{{P_I}}}} \right] \hfill \\
\end{gathered} ,\tag{23}\]
where ${\lambda _{\max }}{\rm{(}}{\bm{H}}_{00}^H{{\bm{H}}_{00}}{\rm{)}}$ is the maximum eigenvalue operation for the matrix ${\bm{H}}_{00}^H{{\bm{H}}_{00}}$, ${R_0}$ is the Euclidean distance between the BS $B{S_0}$ and the MS $M{S_0}$, ${w_0}$ is the shadowing fading over sub-channels between the BS $B{S_0}$ and the MS $M{S_0}$, $\left| {{z_{0,k,n}}} \right|$ is the Nakagami fading over the sub-channel between the $n$th antenna of the BS $B{S_0}$ and the $k$th antenna of the MS $M{S_0}$. Let $\tau {\rm{ = }}{{\frac{{{P_0}}}{{{N_t}}}\frac{{{H_{\rm{0}}}}}{{{\rm{R}}_0^\sigma }}} \mathord{\left/
 {\vphantom {{\frac{{{P_0}}}{{{N_t}}}\frac{{{H_{\rm{0}}}}}{{{\rm{R}}_0^\sigma }}} {{P_I}}}} \right.
 \kern-\nulldelimiterspace} {{P_I}}}$, the relationship between achievable rate and traffic rate of $M{S_0}$ is regarded as
\[{B_W}{\log _2}(1 + \tau ) = \rho \left( {{x_{M0}}} \right).\tag{24}\]
Based on the PDF of traffic rate in equation~\eqref{20}, the PDF of $\tau $ is derived as
\[\begin{gathered}
  {f_\tau }({\text{z}}) = \frac{{\theta {\rho _{{{\min }^\theta }}}B_W^{ - \theta }}}{{\ln 2\cdot(1 + {\text{z}})}}{\left( {{{\log }_2}(1 + {\text{z}})} \right)^{ - \theta  - 1}}
  \left( {{\text{z}} > {{\text{z}}_0} = {2^{{\rho _{\min }}/{B_W}}} - 1} \right) \hfill \\
\end{gathered} .\tag{25}\label{25}\]

Consider that the MS communicates with the closest BS in a PVT cell. Furthermore the probability of the Euclidean distance between an MS and the $i$th near BS can be expressed as
\[\begin{gathered}
  \Pr \left( {i - 1\;{\text{BSs}}\;{\text{in}}\;{\text{a}}\;{\text{circle}}\;{\text{area}}\;{\text{with}}\;{\text{radius}}\;R} \right)
   = \frac{{{{\left( {{\lambda _B}\pi {R^2}} \right)}^{i - 1}}}}{{\left( {i - 1} \right)!}}{e^{ - {\lambda _B}\pi {R^2}}} \hfill \\
\end{gathered} ,\tag{26}\]
where $\Pr \left( {\cdot} \right)$ is the probability operation. Thus, the PDF of the distance $R$ is derived by
\[\begin{gathered}
  {f_{{R_0}}}(R) = \frac{{d\Pr \{ {R_0} \leqslant R\} }}{{dR}}
   = {-}\frac{{d\Pr \{ {R_0} > R\} }}{{dR}}
   = {-}\frac{{{\text{d}}\left( {\frac{{{{\left( {{\lambda _B}\pi {R^2}} \right)}^{\text{0}}}{e^{ - {\lambda _B}\pi {R^2}}}}}{{{\text{0}\text{!}}}}} \right)}}{{{\text{dR}}}}
   = 2\pi {\lambda _B}R{e^{ - \pi {\lambda _B}{R^2}}} \hfill \\
\end{gathered} ,\tag{27}\]
where $\Pr \{ {R_0}{\rm{ > }}R\} $ is the probability that there is not a BS in the circular area with the center ${x_{M0}}$ and the radius $R$. Considering the path loss effect on the distance $R$, the corresponding PDF is derived as
\[{f_{{R_0}^\sigma }}(R) = \frac{1}{\sigma }{R^{\frac{2}{\sigma } - 1}} \cdot 2\pi {\lambda _B}{e^{ - \pi {\lambda _B}{R^{2/\sigma }}}}.\tag{28}\label{28}\]
Furthermore, the downlink transmission power ${P_0}$ between the BS $B{S_0}$ and the MS $M{S_0}$ is expressed as
\[{P_0}{\rm{ = }}{P_I} \cdot \frac{{{N_t} \cdot R_0^\sigma  \cdot \tau }}{{{H_0}}}.\tag{29}\label{29}\]
Based on equation~\eqref{12}~\eqref{25}~\eqref{28} and~\eqref{29}, the characteristic function of ${P_0}$ is derived by
\[\begin{gathered}
  {\phi _{{P_0}}}\left( \omega  \right) = \int\limits_x {{\phi _{{P_I}}}(\omega x){f_{\frac{{{{\text{N}}_t}R_0^\sigma \tau }}{{{H_0}}}}}\left( x \right)dx}  \hfill \\
   = \int\limits_x {\iint\limits_{y,z} {\frac{{y{\phi _{{P_I}}}(\omega x)}}{{z{N_t}}}{f_\tau }\left( z \right){f_{R_0^\sigma }}\left( {\frac{{xy}}{{z{N_t}}}} \right)}}  {f_{{H_0}}}\left( y \right)dxdydz \hfill \\
   = \iint\limits_{y,z} {\frac{{\pi {\lambda _B}}}{{G\left( \omega  \right){z^{\frac{2}{\sigma }}}{y^{\frac{{ - 2}}{\sigma }}} + \pi {\lambda _B}}}{f_{{H_0}}}\left( y \right){f_\tau }\left( z \right)dydz} \hfill \\
\end{gathered} ,\tag{30a}\]
with\[G\left( \omega  \right) = \delta |\omega {|^{{2 \mathord{\left/
 {\vphantom {2 \sigma }} \right.
 \kern-\nulldelimiterspace} \sigma }}}\left[ {1 - j \cdot {\rm{sign}}(\omega ) \cdot \tan {\textstyle{\pi  \over \sigma }}} \right],\tag{30b}\]
where ${f_{{H_0}}}\left( y \right)$ is the PDF of the channel variable ${H_0} = {w_0}\sum\limits_{n = 1}^{{N_t}} {\sum\limits_{k = 1}^{{N_r}} {{{\left| {{z_{0,k,n}}} \right|}^2}} } $. In this paper, the channel variable ${H_i}$ is \textit{i.i.d.} for all channels. Based on equation~\eqref{7a}, the function ${f_{{H_0}}}\left( y \right)$ is expressed by
\[\begin{gathered}
  {f_{{H_0}}}\left( y \right) = \frac{{2{{\left( {\frac{{m\lambda }}{\Omega }} \right)}^{\frac{{{N_t}{N_r}m + \lambda }}{2}}}}}{{\Gamma \left( {{N_t}{N_r}m} \right)\Gamma \left( \lambda  \right)}}{y^{\frac{{{N_t}{N_r}m + \lambda  - 2}}{2}}} {K_{\lambda  - {N_t}{N_r}m}}\left( {2\sqrt {\frac{{m\lambda y}}{\Omega }} } \right)\left( {y > 0} \right) \hfill \\
\end{gathered} .\tag{31}\]

Assume that BS transmission power is dynamically adjusted to meet the required traffic rates for all MSs in the typical PVT cell ${C_0}$. The required BS transmission power for all MSs in ${C_0}$ is expressed by
 \[{\mathbb{P}_{{C_0}}}\mathop  = \limits^{def} \sum\limits_{{x_{Ms}} \in {\prod _M}} {{P_s} \cdot {\bf{1}}\left\{ {{x_{Ms}} \in {C_0}} \right\}} ,\tag{32}\]
where ${P_s}$ is the consumed power transmitted from the BS $B{S_0}$ to the MS $M{S_s}$ in the typical cell ${C_0}$, $\bf{1}\left\{ {...} \right\}$ is an indicator function for gathering together all MSs belong to the same typical PVT cell ${C_0}$. Assume that ${P_s}$ is a series of \textit{i.i.d.} random variables, of which the PDF and the characteristic function of which are denoted as ${f_P}\left( p \right)$ and ${\phi _P}\left( \omega  \right)$, respectively. Based on the Campbell theory in \cite{Stoyan96}, the characteristic function of required BS transmission power ${P_{{C_0}}}$ in the typical PVT cell ${C_0}$ is derived as
\[\begin{gathered}
  {\phi _{{P_{{C_0}}}}}(\omega ) =
  \exp \left[ {\iint\limits_{x,p} {\left( {{e^{j\omega p}} - 1} \right){f_P}(p){\mathbf{1}}\left\{ {x \in {C_0}} \right\}2\pi {\lambda _M}xdpdx}} \right] \hfill \\
   = \exp \left[ { - 2\pi {\lambda _M}\int\limits_0^\infty  {\left( {1 - {\phi _P}(\omega )} \right)\left( {{\mathbf{1}}\left\{ {x \in {C_0}} \right\}} \right)xdx} } \right] \hfill \\
   = \exp \left[ { - 2\pi {\lambda _M}\int\limits_0^\infty  {\left( {1 - {\phi _P}(\omega )} \right){e^{ - \pi {\lambda _B}{x^2}}}xdx} } \right] \hfill \\
   = \exp \left[ { - \frac{{{\lambda _M}}}{{{\lambda _B}}}\left( {1 - {\phi _{{P_0}}}(\omega )} \right)} \right] \hfill \\
\end{gathered} .\tag{33}\label{33}\]
And ${f_{{P_{_{{C_0}}}}}}\left( x \right)$ is the PDF of the required BS transmission power ${P_{{C_0}}}$, which can be calculated by applying the inverse Fourier operation to ${\phi _{{P_{{C_0}}}}}(\omega )$. Considering the limit of BS transmission power ${P_{max} }$~\cite{xiang13}, the energy efficiency of MIMO random cellular networks with the average power allocation scheme is derived by
\[\begin{gathered}
  EE = \frac{{\frac{{{\lambda _M}\theta {\rho _{\min }}}}{{{\lambda _B}(\theta  - 1)}} \cdot {{\left( {\int\limits_0^{{P_{\max }}} {{f_{{P_{{C_0}}}}}\left( x \right)dx} } \right)}^2}}}{{\frac{1}{\eta }\int\limits_0^{{P_{{C_0}\_\max }}} {x{f_{{P_{{C_0}}}}}\left( x \right)dx}  + \left( {{N_T}{P_{dyn}} + {P_{sta}}} \right) \cdot \int\limits_0^{{P_{\max }}} {{f_{{P_{{C_0}}}}}\left( x \right)dx} }} \hfill \\
\end{gathered} .\tag{34}\]

\subsection{Energy Efficiency of MIMO Cellular Networks Using Water-filling Power Allocation Scheme}
When perfect CSI is assumed to be available at both transmitters and receivers in wireless communication systems, the water-filling power allocation scheme is used for improve the capacity of wireless communication systems~\cite{Khoshnevisan12}. The downlink capacity over wireless channels between the BS $B{S_0}$ and the MS $M{S_0}$ is expressed as
\[\begin{array}{l}
\mathbb{C} = \max \log \det \left( {{I_{{N_t}}} + \frac{{{\bm{\varLambda }}_0^H{\bm{D}}_{00}^2{{\bm{\varLambda }}_0}}}{{{P_I}}}} \right)\;\;{\text{subject}}\;{\text{to}}\;\sum\limits_{l = 1}^{{N_t}} {{P_l}}  = {P_{T0}}
\end{array},\tag{35}\label{35}\]
where ${N_0}$ is the noise power at the transmitters. The objective function in ~\eqref{35} is jointly concave in the powers and this optimization problem can be solved by Lagrangian methods\cite{Lu13}, the optimal transmission power for the $l$th sub-channel is given by
\[{P_{l}} = {\left( {\nu  - \frac{{{N_0}}}{{{\lambda _l}}}} \right)_ + } = \frac{{{P_{T0}}}}{L} + \frac{1}{L}\sum\limits_{l = 1}^L {\frac{{{N_0}}}{{{\lambda _l}}}}  - \frac{{{N_0}}}{{{\lambda _l}}},\tag{36}\label{36}\]
where $\nu $ is the water-filling threshold in water-filling power allocation scheme. Based on equation~\eqref{36} and~\eqref{16}, the achievable rate with the water-filling power scheme at the MS $M{S_0}$ is derived as
\[\begin{gathered}
  {\mathcal{R}_0} = {B_W}\log \left[ {1 + \frac{{{\bm{\varLambda }}_0^H{\bm{D}}_{00}^2{{\bm{\varLambda }}_0}}}{{{P_I}}}} \right] = {B_W}\log \det \left( {{I_{{N_t}}} + \frac{{{\bm{\varLambda }}_0^H{\bm{D}}_{00}^2{{\bm{\varLambda }}_0}}}{{{P_I}}}} \right) \hfill \\
  \quad \; = {B_W}\sum\limits_{l = 1}^L {\log \left( {1 + \frac{{{P_l}{\lambda _l}}}{{{P_I}}}} \right)}  = {B_W}\sum\limits_{l = 1}^L {\log \left( {1 + \frac{{\left( {\frac{{{P_{T0}}}}{L} + \frac{1}{L}\sum\limits_{l = 1}^L {\frac{{{N_0}}}{{{\lambda _l}}}}  - \frac{{{N_0}}}{{{\lambda _l}}}} \right){\lambda _l}}}{{{P_I}}}} \right)}  \hfill \\
\end{gathered} .\tag{37}\]
Assume that the traffic rate is satisfied by the achievable rate at the MS $M{S_0}$, thus the corresponding balance equation is expressed by
\[{B_W}\sum\limits_{l = 1}^L {\log \left( {1 + \frac{{\left( {\frac{{{P_{T0}}}}{L} + \frac{1}{L}\sum\limits_{l = 1}^L {\frac{{{N_0}}}{{{\lambda _l}}}}  - \frac{{{N_0}}}{{{\lambda _l}}}} \right){\lambda _l}}}{{{P_I}}}} \right)}  = \rho \left( {{x_{M0}}} \right){\text{ }} .\tag{38}\label{38}\]
When the maximal BS transmission power limit ${P_{\max }}$ is considered, based on the equation~\eqref{38} the Monte-Carlo simulation is configured to iteratively solve the transmit power ${P_{T0}}$. Then, $\mathbb{E}\left\{{{P_{{C_0}\_real}}}\right\}$ and ${F_{{P_{{C_0}}}}}\left( {{P_{\max }}} \right)$ can be averaged and statistically computed from the simulation results. Substituting the values for $\mathbb{E}\left\{{{P_{{C_0}\_real}}}\right\}$ and ${F_{{P_{{C_0}}}}}\left( {{P_{\max }}} \right)$ into equation~\eqref{19}, the energy efficiency of MIMO PVT cellular networks with the water-filling power allocation scheme can be obtained as
\[EE = \frac{{\frac{{{\lambda _M}\theta {\rho _{\min }}}}{{{\lambda _B}(\theta  - 1)}}{{\left( {{F_{{P_{{C_0}}}}}\left( {{P_{\max }}} \right)} \right)}^2}}}{{\frac{1}{\eta }\mathbb{E}\left\{ {{P_{{C_0}\_real}}} \right\} + \left( {{N_T}{P_{dyn}} + {P_{sta}}} \right) \cdot {F_{{P_{{C_0}}}}}\left( {{P_{\max }}} \right)}}.\tag{39}\]

\begin{figure}[!t]
\centering
\begin{minipage}[c]{0.48\textwidth}
\centering
\includegraphics[width=7cm,draft=false]{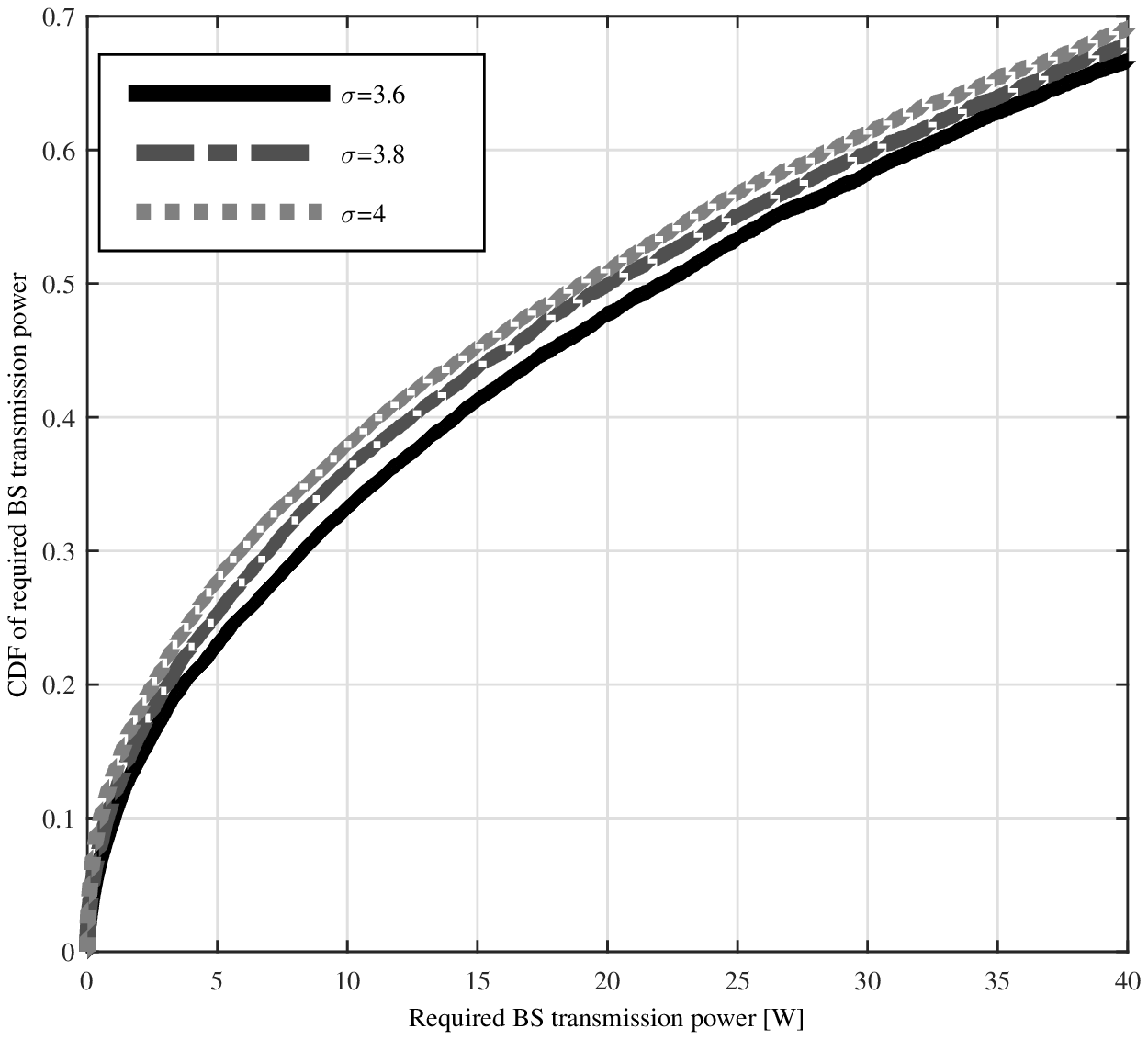}
\end{minipage}
\hspace{0.02\textwidth}
\begin{minipage}[c]{0.48\textwidth}
\centering
\includegraphics[width=7cm,draft=false]{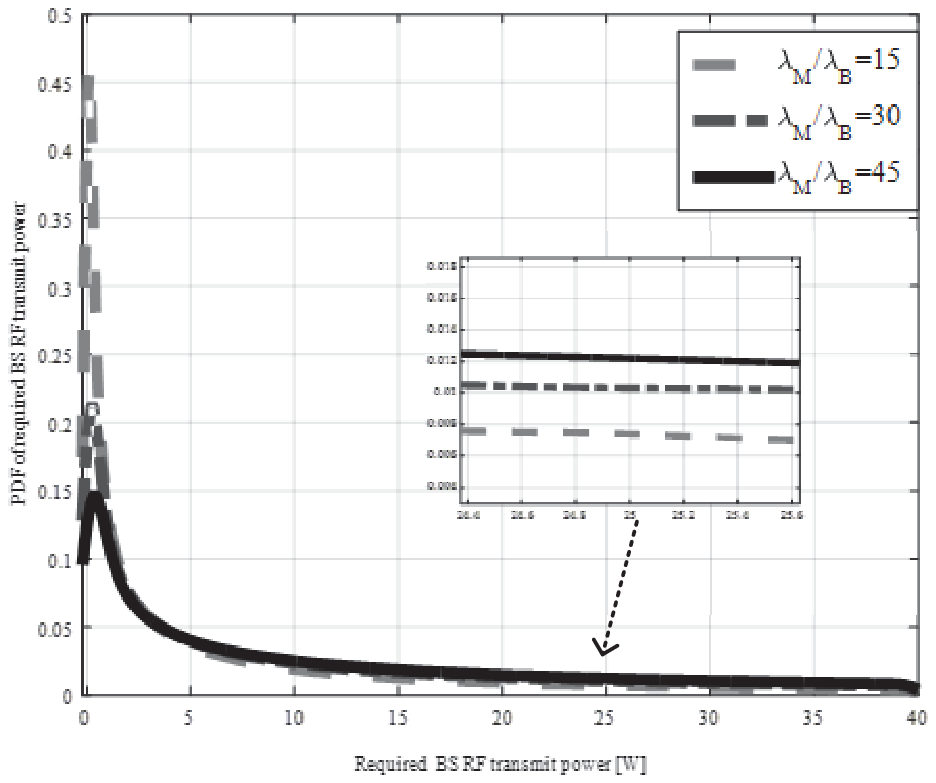}
\end{minipage}\\[1mm]
\begin{minipage}[t]{0.48\textwidth}
\centering
\caption{The CDF of the required BS transmission power with water-filling power allocation scheme.}
\label{F4}
\end{minipage}
\hspace{0.02\textwidth}
\begin{minipage}[t]{0.48\textwidth}
\centering
\caption{The PDF of the required BS transmission power with water-filling power allocation scheme.}
\label{F5}
\end{minipage}
\end{figure}

Based on equation~\eqref{38}, the CDF and PDF of the required BS transmission power with the water-filling power allocation scheme are analyzed as follows. Unless otherwise specified, the key parameters are set as ${\sigma _{dB}}{\rm{ = }}6$, $\sigma{\rm{ = }}4$, $m{\rm{ = }}1$ \cite{Simon00}, ${N_t}{\rm{ = }}8$, ${N_r}{\rm{ = 4}}$, ${P_{\max }}{\rm{ = 40}}$ Watt (W) \cite{Arnold10}, the moment of receiving power $\mathbb{E}\left\{{P_{Ti}^{2/\sigma }}\right\}{\rm{ = }}{10^{{\rm{ - }}2}}$ W, ${\lambda _B}{\rm{ = }}1/\left( {\pi  \cdot {{800}^2}} \right){\rm{ }}{{\rm{m}}^{ - 2}}$, ${\lambda _M}/{\lambda _B}{\rm{ = }}30$, ${\lambda _{\inf }}{\rm{ = }}0.9{\lambda _B}$, $\theta {\rm{ = }}1.8$, ${\rho _{\min }}{\rm{ = }}2.5\;{\rm{bits/s/Hz}}$ \cite{xiang13}. Figure~\ref{F4} reveals the the CDF of the required BS transmission power with water-filling power allocation scheme considering different path loss coefficients $\sigma$. Figure~\ref{F4} indicates that the CDF curve shifts to the left with the increasing value of $\sigma$, i.e., the required BS transmission power with the water-filling power allocation scheme is decreased when the value of $\sigma$ is increased in MIMO PVT random cellular networks.

Figure~\ref{F5} evaluates the impact of intensity ratio of MSs to BSs ${\lambda _M}/{\lambda _B}$ on the required BS transmission power with the water-filling power allocation scheme. Figure~\ref{F5} shows that the probability mass shifts to the right when increasing the value of ${\lambda _M}/{\lambda _B}$, i.e., the required BS transmission power with water-filling power allocation scheme is increased when the value of ${\lambda _M}/{\lambda _B}$ is higher.

\section{Performance Analysis and Discussion}
The effect of two power allocation schemes on the proposed energy efficiency model of MIMO random cellular networks is investigated in detail following. In the following simulations, the Monte-Carlo simulation method is adopted for performance analysis. Moreover, the total BS transmission power including the required BS transmission power and RF circuit power, the BS fixed power is investigated in this section. Default system parameters are configured as: the average efficiency of power amplifier is $\eta {\rm{ = }}0.38$, the RF circuit power for an antenna  ${P_{dyn}} = 83$W and the fixed power consumption in a BS is ${P_{sta}} = 45.5$W~\cite{Xu13,Hong13}.

\begin{figure}[!t]
\centering
\begin{minipage}[c]{0.48\textwidth}
\centering
\includegraphics[width=8cm,draft=false]{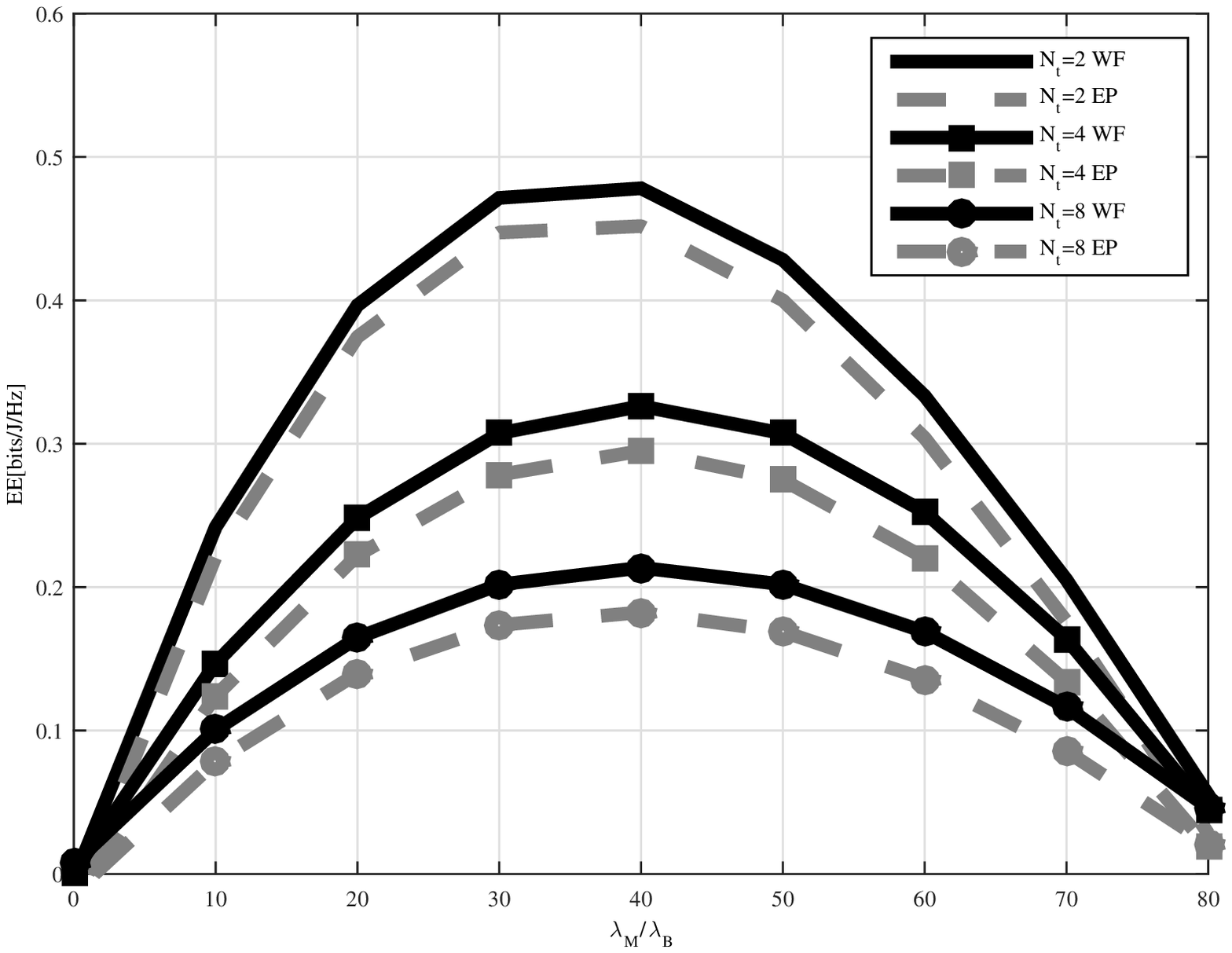}
\end{minipage}
\hspace{0.02\textwidth}
\begin{minipage}[c]{0.48\textwidth}
\centering
\includegraphics[width=8cm,draft=false]{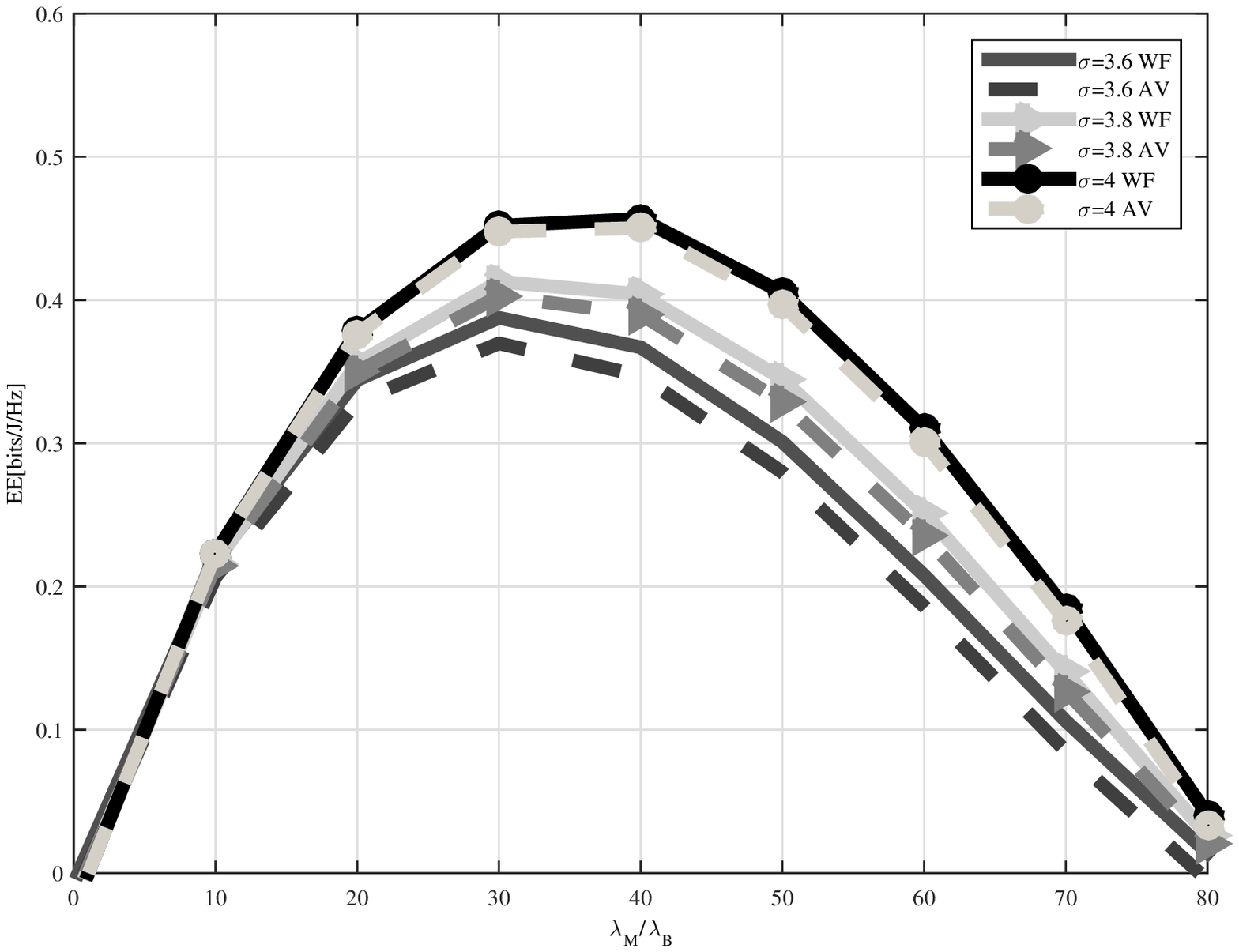}
\end{minipage}\\[1mm]
\begin{minipage}[t]{0.48\textwidth}
\centering
\caption{Energy efficiency of MIMO random cellular networks versus ${\lambda _M}/{\lambda _B}$ and ${N_t}$.}
\label{F6}
\end{minipage}
\hspace{0.02\textwidth}
\begin{minipage}[t]{0.48\textwidth}
\centering
\caption{Energy efficiency of MIMO random cellular networks versus ${\lambda _M}/{\lambda _B}$ and $\sigma$.}
\label{F7}
\end{minipage}
\end{figure}

 Figure~\ref{F6} illustrates the energy efficiency of MIMO random cellular networks versus the number of transmitting antennas ${N_t}$ and the intensity ratio of MSs to BSs ${\lambda _M}/{\lambda _B}$, where ``WF" denotes the water-filling power allocation scheme and ``AV" represents the average power allocation scheme. First, Figure~\ref{F6} shows that the energy efficiency curve of MIMO random cellular networks shrinks down when increasing the value of ${N_t}$. Base on the result of ~\eqref{22}, the total BS power consumption increases with the increase of the number of transmit antennas, while the average traffic rate remains unchanged. Hence, the energy efficiency of PVT random cellular networks decreases with the increase of ${N_t}$. Second, we force on one of curves and analyze the energy efficiency for both power allocation schemes with impact of ${\lambda _M}/{\lambda _B}$. Simulation results indicate that the water-filling/average power allocation schemes can achieve the maximum energy efficiency for MIMO random cellular networks. When the intensity ratio of MSs to BSs is low, indicating a few MSs in a typical PVT cell, the increase of the intensity ratio of MSs to BSs conduces to a moderate increase in total BS power consumption including mainly fixed BS power consumption and a small portion of dynamic BS power consumption. In this case, the energy efficiency of PVT cellular networks is increased. However, when the intensity ratio of MSs to BSs in a PVT typical cell exceeds a given threshold, a high aggregate traffic load resulted from a large number of MSs will significantly increase the total BS power consumption including mainly dynamic BS power consumption and a small portion of fixed BS power consumption. In this case, the energy efficiency of PVT cellular networks is decreased. Moreover, the energy efficiency of the water-filling power allocation scheme is always larger than for the energy efficiency of the average power allocation scheme in MIMO random cellular networks.

 Figure~\ref{F7} reveals the impact of the path loss coefficient $\sigma$ and ${\lambda _M}/{\lambda _B}$ on the energy efficiency of MIMO random cellular networks. The energy efficiency curve lifts up as the value of ${\lambda _M}/{\lambda _B}$ increases. Again the energy efficiency of the water-filling power allocation scheme is always larger than for average power allocation scheme under different values of $\sigma$.

\begin{figure}[!t]
\centering
\begin{minipage}[c]{0.48\textwidth}
\centering
\includegraphics[width=8cm,draft=false]{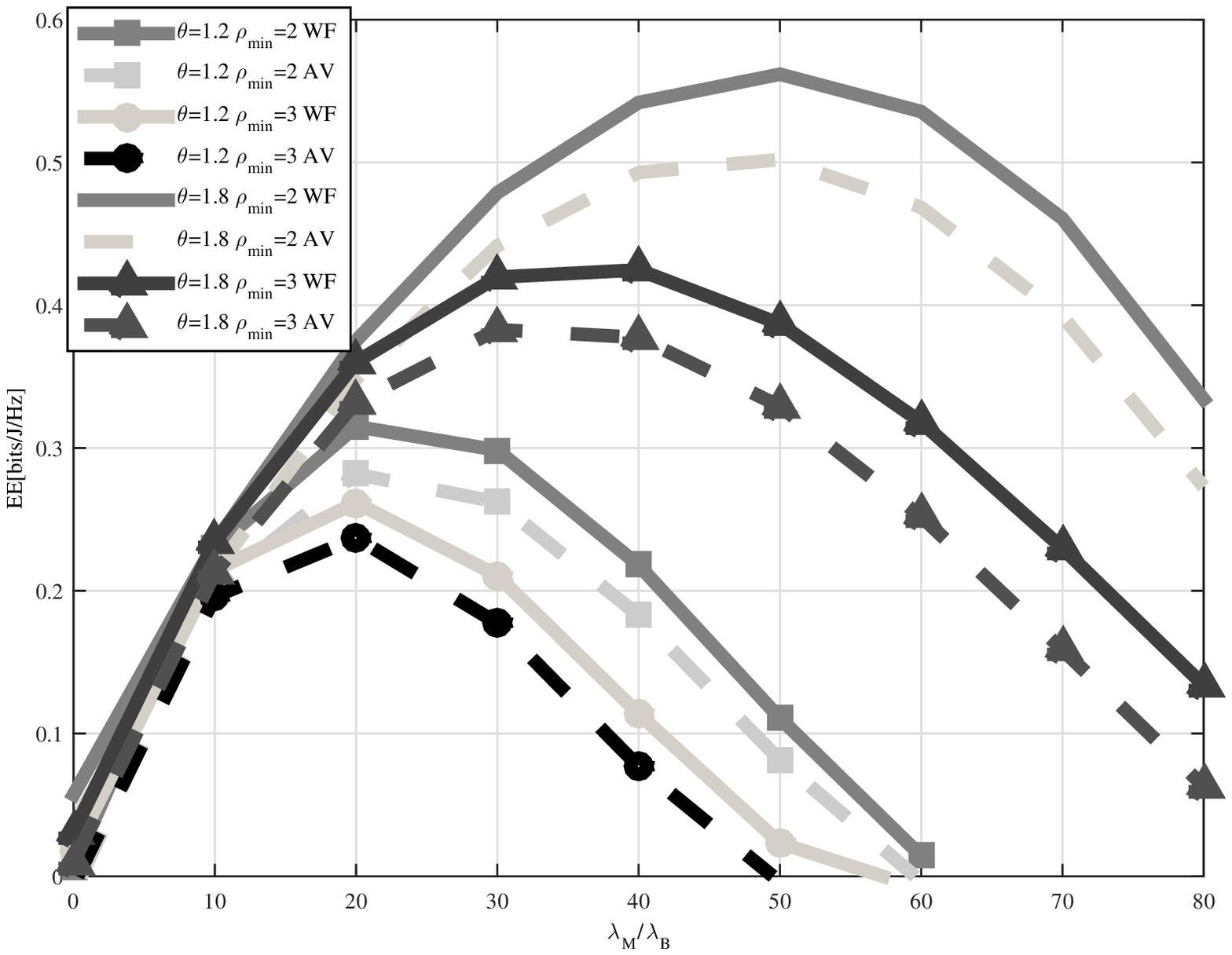}
\end{minipage}
\hspace{0.02\textwidth}
\begin{minipage}[c]{0.48\textwidth}
\centering
\includegraphics[width=8cm,draft=false]{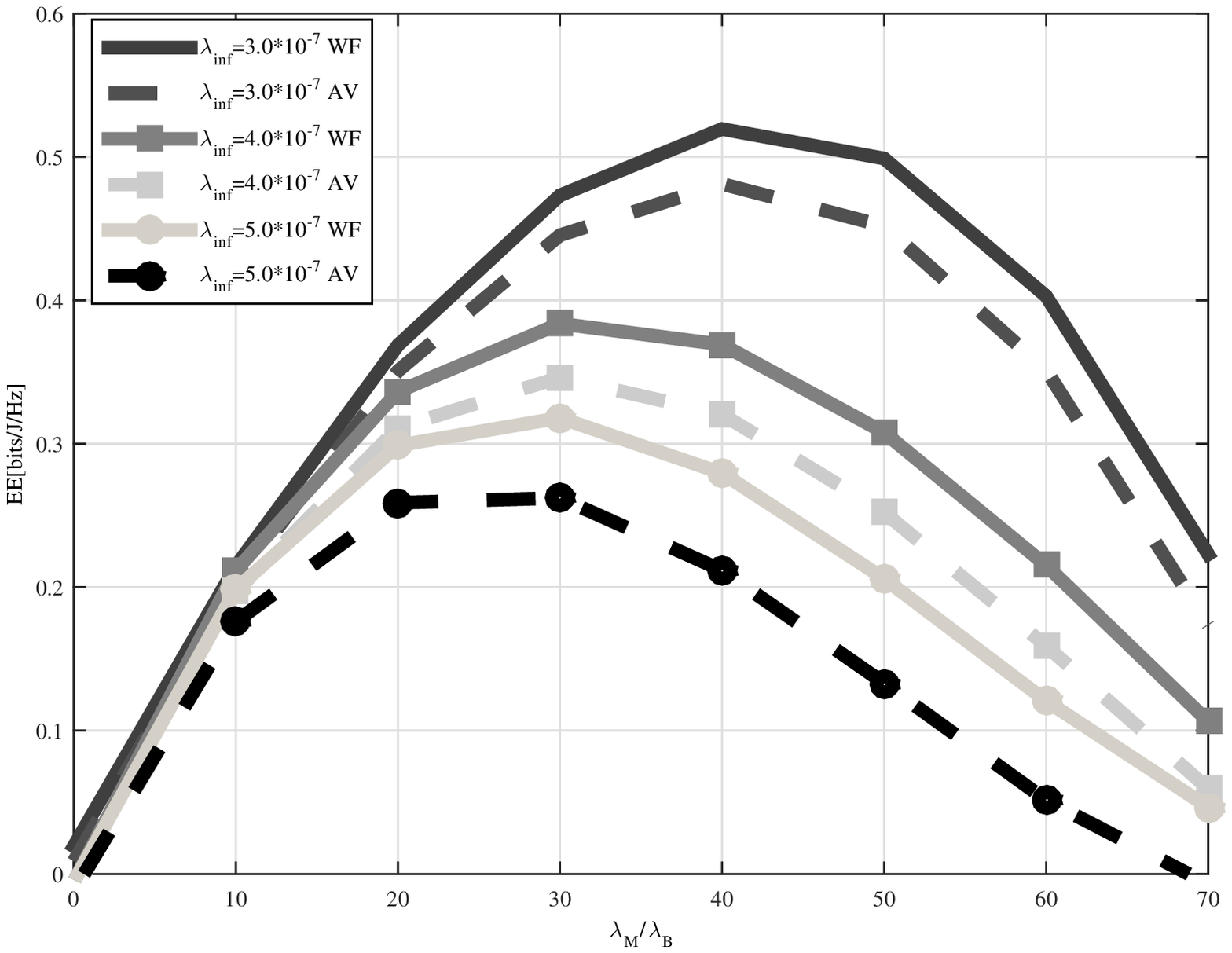}
\end{minipage}\\[1mm]
\begin{minipage}[t]{0.48\textwidth}
\centering
\caption{Energy efficiency of MIMO random cellular networks versus ${\lambda _M}/{\lambda _B}$, $\theta $ and ${\rho _{\min }}$.}
\label{F8}
\end{minipage}
\hspace{0.02\textwidth}
\begin{minipage}[t]{0.48\textwidth}
\centering
\caption{Energy efficiency of MIMO random cellular networks versus ${\lambda _M}/{\lambda _B}$ and $\lambda _{\inf }$.}
\label{F9}
\end{minipage}
\end{figure}

 Figure~\ref{F8} evaluates the effect of the minimum traffic rate ${\rho _{\min }}$ and the tail index $\theta $ on the energy efficiency of MIMO random cellular networks with the two power allocation schemes. There always exists a maximum energy efficiency of MIMO random cellular networks considering different system parameters. However, numerical results indicate that the energy efficiency of the water-filling power allocation scheme is always larger than for the energy efficiency of average power allocation scheme in MIMO random cellular networks.

 Finally, the effect of the interfering BS intensity $\lambda _{\inf }$ on the energy efficiency with different power allocation schemes is investigated in Figure~\ref{F9}. When the values of ${\lambda _M}/{\lambda _B}$ is fixed, the energy efficiency decreases when the value of $\lambda _{\inf }$ increases. The curves in Figure~\ref{F9} indicate that the energy efficiency of the water-filling power allocation scheme is always larger than for the energy efficiency of average power allocation scheme under different values of $\lambda _{\inf }$ in MIMO random cellular networks.The reason is that the water-filling power allocation scheme substantially reduces the sum power, by up to 80\%, in comparison to the average power allocation scheme~\cite{Chen07}. The water-filling power allocation efficiently exploits the multiuser MIMO channels (e.g., multiuser diversity), hence power reduction is more significant. So the water-filling power allocation scheme is better than the average power allocation scheme in the energy efficiency of cellular networks when the traffic rate of MSs is same.

\section{Conclusions}
In this paper, the energy efficiency of MIMO random cellular networks with different power allocation schemes are evaluated. Considering the path loss, Nakagami-m fading and shadowing effects on wireless channels, an interference model and the achievable rate of MIMO random cellular networks is first presented. Furthermore, the energy efficiency of average and water-filling power allocation schemes is proposed, respectively. Simulation results indicate that there exists a maximal network energy efficiency when considering the trade-off between intensity ratios of MSs to BSs and wireless channel conditions. When the CSI is available for both transmitters and receivers, the energy efficiency of the water-filling power allocation scheme is better than the energy efficiency of average power allocation scheme in MIMO random cellular networks. Therefore, our results evaluate the impact of different power allocation schemes on the energy efficiency of MIMO random cellular networks. For the future work, we will use obtained results to analyze future 5G heterogeneous network adopting the millimeter wave transmission technology.

\Acknowledgements{This work was supported by the National Natural Science Foundation of China (Grant Nos. 60872007, 61271224);
the Major International (Regional) Joint Research Program of China (Grant No. 61210002);
the Ministry of Science and Technology of China through the 863 Project in 5G under
Grant 2014AA01A701;
the Ministry of Science and Technology (MOST) of China (Grant No. 2015FDG12580);
the Hubei Provincial Science and Technology Department (Grant No. 2013BHE005);
the Fundamental Research Funds for the Central Universities (Grant Nos. 2015XJGH011, 2013ZZGH009); the EU FP7 QUICK project under Grant PIRSES-GA-2013-612652; 
EU H2020 5G Wireless project under Grant 641985;
EU FP7-PEOPLE-IRSES, project acronym S2EuNet (Grant no. 247083), project acronym
WiNDOW (grant no. 318992) and project acronym CROWN (grant no. 610524);
UK EPSRC Grant EP/J015180/1;
Ms. Peipei Song involved some initial discussions for this paper and we thank her useful suggestions on this paper.}

\InterestConflict





\end{document}